\documentclass{article}
\usepackage{amsmath}
\usepackage{amssymb}
\usepackage{graphicx}
\usepackage{bm}
\title{Self-stresses and Crack Formation by Particle Swelling in Cohesive Granular Media}
\author{M. S. El Youssoufi\footnote{Correspondance to: M.S. El Youssoufi (elyous@lmgc.univ-montp2.fr)}, J.-Y. Delenne and F. Radjai\\
LMGC, CNRS-Universit\'e Montpellier II,\\ Place Eug\`ene Bataillon,\\ 34095 Montpellier cedex, France.}

\date{\today}

\begin{document}
\maketitle
\begin{abstract}
We present a molecular dynamics study    
of force patterns, tensile strength and crack formation in a cohesive granular model  
where the particles are subjected to swelling or shrinkage gradients.  
Non-uniform particle size change generates self-equilibrated forces 
that lead to crack initiation as soon as strongest tensile contacts begin to fail. 
We find that the coarse-grained stresses are correctly predicted by 
an elastic model that incorporates 
particle size change as metric evolution. The tensile strength is found to be 
well below the theoretical strength as a result of inhomogeneous 
force transmission in granular media. The cracks propagate either inward from 
the edge upon shrinkage and outward from the center upon swelling.
\end{abstract}

\noindent 83.80.Fg, 74.80.-g, 45.70.Mg\\

\noindent \textbf{key words:} granular media / porous media / cohesion / shrinkage / swelling / rupture\\
\vspace{0.5cm}


The term ``cohesive granular media'' covers a vast 
spectrum of granular materials in which 
rigid grains are bound together by cohesion forces of various 
chemico-physical origins \cite{maugis99}. Well-known examples 
are fine powders and 
soils with more or less colloidal or water content \cite{mitchell93}. 
The solid-like behavior attributed to 
noncohesive granular media under quasistatic shearing \cite{jaeger96b} becomes 
the dominant feature in the presence of cohesion,  
with an increasing effective tensile strength as a function of the contact 
tensile strength \cite{kim03}. The stress-stain behavior and 
fracture mechanics of cohesive granular media raise interesting open 
issues from a grain-scale point of view and  
in interaction with heat or mass transfer.       

An appealing issue is 
how and in which respects these ``granular solids'' differ from 
molecular solids (in the absence of a granular structure).  
For example, the phenomenon of 
stress concentration, induced by defects at different scales,  
governs the initiation of failure 
in molecular solids, the effective tensile strength remaining generally far below the 
``theoretical'' strength \cite{herrmann90}.  In a granular assembly, 
stress concentration occurs already at the particle scale in the form 
of a highly inhomogeneous distribution of contact forces \cite{mueth98,radjai98b}. 
This suggests that,  
even in the absence of mesoscopic defects, the tensile strength 
will be weak compared to its theoretical value  
for a granular assembly (to be defined below). 
However, the tensile-strength properties  
have scarcely been analyzed from a microscopic 
standpoint \cite{kim03,radjai00a}.

\begin{figure}
	\begin{center}
	  \includegraphics[width=0.8\textwidth]{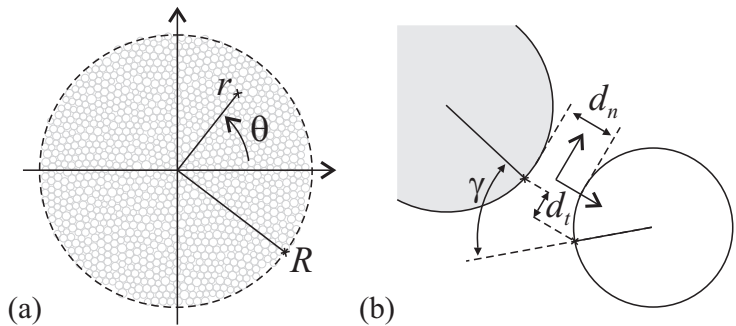}
	  \caption{(a) Geometry of the sample; (b) Relative displacements 
                   between two edge points belonging to  
                   two particles and coinciding initially with their contact point.}
	 \label{fig1}
	\end{center}
\end{figure}

In this Letter, we consider a benchmark test that was designed to 
probe the {\em intrinsic} tensile response (reflecting only 
the granular disorder) of a cohesive granular sample 
by avoiding both wall effects and strain localization as spurious sources 
of randomness.  
The sample consists of rigid cohesive disks  compacted 
numerically into a circular form  in a two-dimensional space; See Fig. \ref{fig1}(a).  
At the start, the normal force is exactly zero at all contacts. 
Then, the particle diameters are increased (or decreased) at a rate that depends on 
distance to the sample center. 
Such gradients of particle size change 
occur, for instance, in fine soils, where  
particle swelling (or shrinkage) happens as a result of 
humidification (or drying) \cite{mitchell93}. 
As we shall see in detail below, 
this bulk straining induces 
a field of radial (or orthoradial) tensile self-stresses increasing 
in magnitude with time, and leading 
eventually to crack initiation at the center (or on the edge).

For the simulations, we used the molecular dynamics  method with a 
velocity-Verlet scheme for the integration of the equations 
of motion \cite{allen87}.  Cohesive interaction between two particles implies resistance to 
relative motion (normal displacement $d_n$, tangential displacement $d_t$ and 
angular displacement $\gamma$) of two edge points belonging respectively to the 
two particles and coinciding initially with the contact point; Fig. \ref{fig1}(b).     
The corresponding contact actions are the normal force $f_n$, the tangential 
force $f_t$ and the contact torque $M$. We assume a linear elastic behavior 
characterized by three stiffnesses $E_n$, $E_t$ and $E_\gamma$, so that 
$f_n=E_n d_n$,  $f_t=E_t d_t$ and $M=E_\gamma \gamma$. As usual, damping 
actions are added in order to account for  
contact inelasticity and ensure numerical stability. 

\begin{figure}
	\begin{center}
	  \includegraphics[width=0.8\textwidth]{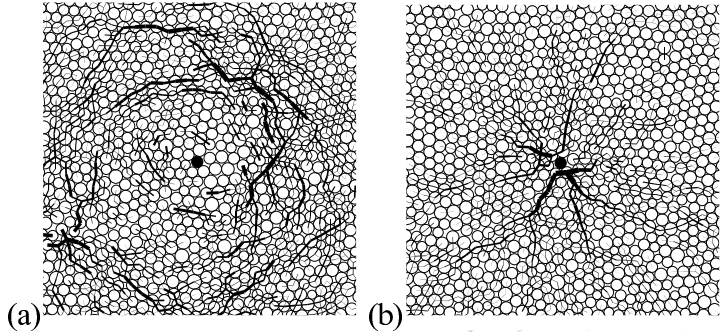}
	  \caption{Tensile (a) and compressive (b) normal forces generated by the swelling of 
                   a single particle (in black). The line width is proportional to 
                   the normal force.}
    \label{fig2}
	\end{center}
\end{figure}

This elastic behavior 
holds as long as the contact actions remain below a ``yield surface'' 
$\zeta = \zeta(f_n, f_t, M)=0$. We used the following function 
that fits our previous experimental tests with a particular type of glue used  
to stick cylindrical particles together \cite{delenne02}: 
\begin{equation}
\zeta=\left(  \frac{f_n}{f_n^y}\right)  
+\left(  \frac{f_t}{f_t^y}  \right)^2
+\left(  \frac{M}{M^y}    \right)^2  -  1, 
 \label{eqn1}
\end{equation}       
where $f_{n}^y$, $f_{t}^y$ and $M^y$ are the yield parameters for normal, tangential 
and angular actions, respectively. The elastic domain corresponds to $\zeta <0$. Note that 
$f_n$ can take indefinitely large values (the positive sign corresponding to 
compressive forces) but it has a lower bound $f_n= - f_n^y$ that defines the 
largest tensile force that can be sustained by a contact.  As soon as 
$\zeta \geq 0$, the cohesive bond breaks down irreversibly and the contact turns into 
noncohesive frictional behavior \cite{luding98c}.
    

In general, the shape of the yield function $\zeta$ and the 
values of the parameters will influence the failure properties 
of the material as a whole 
for a specified loading path.  In our system, loading 
by particle swelling or shrinkage induces appreciable displacements only along 
the contact normals. As a result, the behavior is not sensitive to the 
choice of the values of $f_t^y$ and $M^y$, and 
the failure is 
governed by extensional strain when $f_n^y$ is reached at a strongly tensile 
contact in the sample.                
 
 
We used samples composed of 1133 polydisperse disks with a 
uniform distribution of diameters $D$ within 
a range $[D_{min}, D_{max}]$ where $D_{max}=1.2 \ D_{min}$.
The coefficient of friction is $\mu=0.1$. 
Each sample is created by removing all particles 
belonging to a noncohesive assembly in static equilibrium except those  
contained inside a circle of radius $R_0$.  The   
cohesive bonds are then switched on and the sample is allowed to relax to 
equilibrium.  The samples prepared by this procedure correspond to a 
dense but disordered packing of solid fraction $\simeq 0.89$ and 
coordination number $3.8$.   


At the beginning, the system is in its reference state with the contact actions 
being identically zero  everywhere. Obviously, 
multiplying all particle diameters by the same 
factor does not disturb this state since no relative motions 
are generated at the contact points, whereas {\em differential} particle-size change 
gives immediately rise to permanent compressive and tensile force gradients.   
For instance, the swelling of a single particle 
produces compressive radial forces by pushing the neighboring particles 
outward, as well as tensile orthoradial forces by increasing the total length of 
the ``rings'' of contiguous particles surrounding the swelling particle; Fig. \ref{fig2}. 
A slight shrinkage of the same particle produces exactly the same force patterns 
with the signs inverted everywhere 
(compressive contacts turning to tensile, and vice versa). 

Since we are interested here only in the effect of bulk straining, 
we require that the swelling rate ${\dot s}_i \equiv {\dot D_i} / D_i$ of each particle  
$i$ is independent of its diameter $D_i$. We use the simplest swelling kinetics 
defined by a constant gradient from the center to the edge, 
${\dot s}_i = \frac{\alpha}{R_0} \ r_i$, 
where $r_i$ is the distance of the particle to the system center, and $\alpha$ 
is a constant rate. Positive and negative values of $\alpha$ correspond to particle 
swelling and shrinkage, respectively. 

\begin{figure}
	\begin{center}
	  \includegraphics[width=0.8\textwidth]{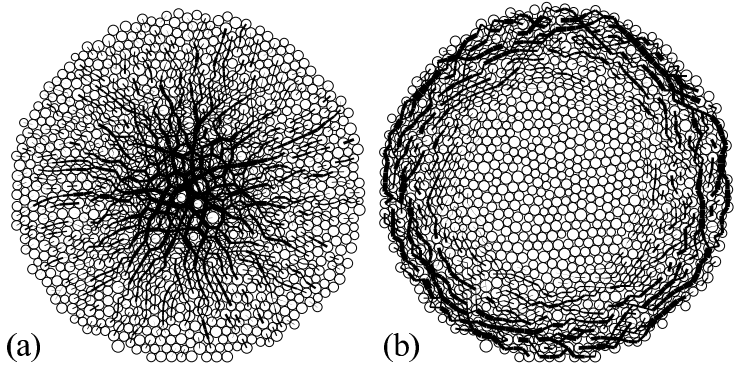}
	  \caption{Compressive (a) and tensile (b) forces in a shrinkage simulation.}
   \label{fig3}
	\end{center}
\end{figure}

\begin{figure}
	\begin{center}
	  \includegraphics[width=0.8\textwidth]{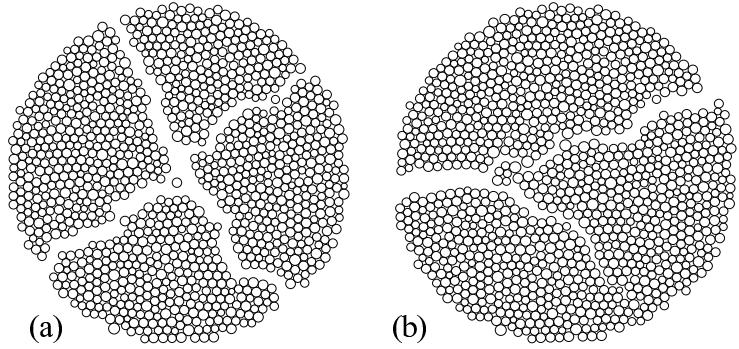}
	  \caption{Crack patterns in swelling (a) and shrinkage (b) simulations.}
	  \label{fig4}
	\end{center}
\end{figure}

Figure \ref{fig3} shows snapshots of normal compressive and tensile 
forces in a shrinkage simulation. 
Although at the very local scale the forces are inhomogeneously distributed, 
we observe radial and orthoradial compressive forces decreasing in magnitude 
from the center to the edge, as well as orthoradial tensile forces decreasing in 
magnitude from the edge to the center. The cracks appear on the edge 
as soon as the first tensile contact fails, and they propagate towards the center as 
shown in Fig. \ref{fig4}(b).
In swelling simulations, the respective 
roles of compressive and tensile roles are simply interchanged with respect to the 
shrinkage case. As a result, the cracks are initiated at the center, and they propagate towards 
the edge; Fig. \ref{fig4}(a).

\begin{figure}
	\begin{center}
	  \includegraphics[width=0.8\textwidth]{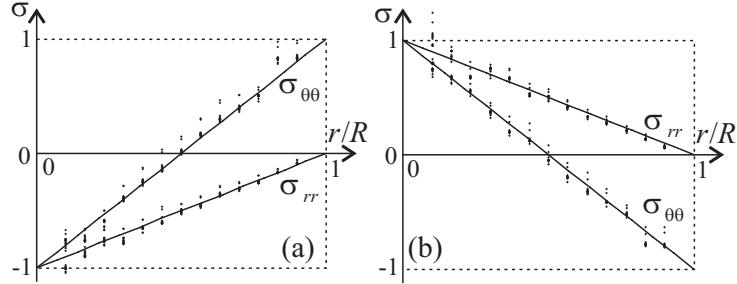}
	  \caption{Normalized stress components 
                   as a function of distance $r$ to the center in swelling (a) and
                   shrinkage (b) simulations. The full lines are elastic fits.}
	  \label{fig5}
	\end{center}
\end{figure}


The coarse-grained stresses can be evaluated from the contact forces 
by means of the ``micromechanical'' expression of the stress 
tensor $\bm \sigma$ \cite{christoffersen81}:
\begin{equation}
\sigma_{ij} = \frac{1}{V} \sum_{c \in V} \  f_i^c \ \ell_j^c,
\label{eqn3}
\end{equation}
where $i$ and $j$ design the coordinates, $V$ is the control volume over which 
the stress tensor is evaluated (the contacts $c$ taken from this volume), 
$f_i^c$ is the $i$ component of the 
force ${\bm f}^c$ at contact $c$, and $\ell_j^c$ is the $j$ component of 
the vector $\bm \ell^c$ joining the centers of the two contact neighbors.

The stress tensor   is a well-defined average if 
the control volume $V$ contains  
a sufficiently large number of contacts. This requirement is satisfied by taking 
concentric circular volume elements as suggested by the rotational invariance of 
our system. We use polar coordinates and the radial and angular positions 
will be denoted by $r$ and $\theta$, respectively; Fig. \ref{fig1}(a).  
As a result of isotropic straining, the cross 
components $\sigma_{r \theta}$ are zero.

The radial stress $\sigma_{r r}$ 
and orthoradial stress $\sigma_{\theta \theta}$ are shown in 
Fig. \ref{fig5}(a) as a function 
of distance $r$ to the center at different  
stages of a swelling simulation. For each data set at a given instant of 
evolution, we have normalized the distance $r$ by $R$ and the stress components by    
the largest tensile stress $-\sigma_{max}$ (occurring at the center).  
We see that the normalized stresses collapse on a straight line 
as a function of $r$ and they nicely agree       
with a one-parameter analytical fit that will be detailed below.  
Note that   $\sigma_{r r}$ is negative (tensile stress) throughout the sample, whereas 
$\sigma_{\theta \theta}$ changes sign at $r \simeq R/2$. In the case of 
shrinkage simulations, similar results are obtained   
with opposite signs, as shown in Fig. \ref{fig5}(b).     

\begin{figure}
	\begin{center}
	  \includegraphics[width=0.8\textwidth]{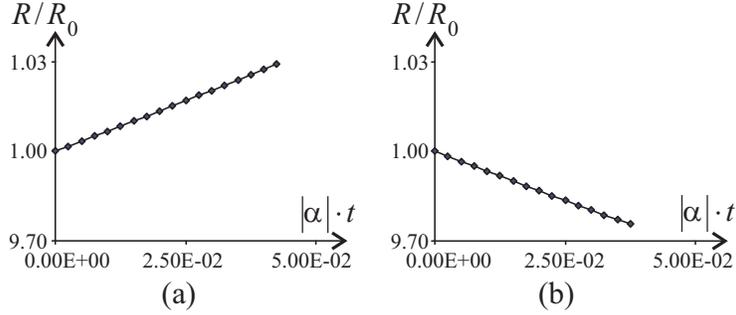}
	  \caption{Evolution of the sample radius $R$ in swelling (a) and 
                   shrinkage (b) simulations.}
	  \label{fig6}
	\end{center}
\end{figure}


We now turn to analytical evaluation of the stresses. 
At a coarse-grained scale, the granular assembly will be represented by a 
linear elastic medium with an effective stiffness $E$ and an effective 
Poisson's ratio $\nu$. This is a plausible assumption although, as 
we shall see below, the behavior of the stresses as a function of $r$ 
is independent of the nature of the interactions.   
On the other hand, coarse-grained swelling 
at a point $A$ of polar coordinates $(r,\theta)$ in space 
is equivalent to an imposed isotropic {\em space dilation} 
${\dot s}(r,\theta) = \langle {\dot s}_i \rangle_{i \in V(r,\theta)}$ , 
where the average is taken 
over all particles contained in a representative volume $V(r,\theta)$ centered 
on $A$. Since the rates are independent of particle diameters, we get  
\begin{equation}
{\dot s} = \frac{\alpha}{R_0} \ r.
\label{eqn4}
\end{equation}         
Then, the strain-rate tensor $\dot {\bm \varepsilon}$  at a point is the sum of two terms: 
\begin{equation}
{\dot {\bm \varepsilon}} = {\dot {\bm \varepsilon}}^e + {\dot s} {\bm I},
\label{eqn5}
\end{equation}
where $ {\dot {\bm \varepsilon}}^e$ is the elastic strain rate, 
$ {\bm I}$ represents the unit tensor, and  ${\dot s} {\bm I}$ is the 
metric change rate.

We assume that the displacement field ${\bm u}(r,\theta)$ is radial (according to the 
symmetry of straining expressed by Eq. \ref{eqn4} and that of the sample) so that 
${\dot \varepsilon}_{r \theta} \equiv 0$. 
Since the radius $R$ of the sample changes with time, 
Hooke's laws will be written in the form of rate equations:
\begin{equation}
\left\{
\begin{array}{lclcl}
{\dot \varepsilon}^e_{rr} & = &  {\dot \varepsilon}_{rr} - {\dot s} 
                                        & = & - \frac{1}{E} ({\dot \sigma}_{rr} 
                                                    - \nu  \ {\dot \sigma}_{\theta\theta}),  \\
{\dot \varepsilon}^e_{\theta \theta} & = &  {\dot \varepsilon}_{\theta\theta} - {\dot s} 
                                        & = & - \frac{1}{E} ({\dot \sigma}_{\theta\theta}  
                                                    - \nu  \ {\dot \sigma}_{rr}),
\end{array}
\right.
\label{eqn6}
\end{equation}
where extensional strains and compressive stresses are counted positive. 
We need also the balance equation which takes the following form in polar coordinates: 
\begin{equation}
{\dot \sigma}_{\theta\theta} -  {\dot \sigma}_{rr} = 
r \ \frac{\partial {\dot \sigma}_{rr}}{\partial r}.
\label{eqn7}
\end{equation}

The set of equations \ref{eqn4}, \ref{eqn6} and \ref{eqn7} is easily integrated 
over time  and space using the boundary 
conditions ${\bm u}(r=0)=0$ (imposed by \ref{eqn4}) 
and $\sigma_{rr} (r=R) = 0$ (by continuity of the normal stress at the 
boundary). The solution is 
\begin{equation}
\left\{
\begin{array}{lcl}
R &=& \frac{R_0}{1 - 2\alpha t/3}, \\
\sigma_{rr} &=& E \left(  1 - \frac{R_0}{R} \right) \left(  \frac{r}{R}  - 1\right), \\ 
\sigma_{\theta\theta} &=&  E \left(  1 - \frac{R_0}{R} \right) \left(  2 \frac{r}{R}  - 1\right). 
\end{array}
\right.
\label{eqn8}
\end{equation}
We see that both stress components are linear in $r$. The simulation 
data of Fig. \ref{fig5} were fitted by adjusting only 
the effective elastic modulus $E$. 
The evolution of the system is, however, nonlinear as a function of time. 
The evolution of $R$ is shown in Fig. \ref{fig6} for swelling  
($\alpha > 0$) and shrinkage ($\alpha < 0$) simulations 
together with the analytical fit 
from Eq. \ref{eqn8} which involves no fitting parameter. The agreement is 
excellent although the nonlinear nature of the evolution can not be seen 
for $|\alpha|t \ll 1$.    
The largest tensile stress $\sigma_{max}$ occurs on the edge for shrinkage 
and at the center for swelling. From Eq. \ref{eqn8}, we get 
$\sigma_{max} = \frac{2}{3} E \ |\alpha| t$.  
Again, this linear form nicely fits the evolution of $\sigma_{max}$ (by virtue 
of the fits shown in Fig. \ref{fig5}) up to failure 
for $\sigma_{max} = \sigma^y$. The latter represents 
the effective tensile strength of the material. 

It is worth noting that, Posisson's ratio $\nu$ does not appear  
in Eqs. \ref{eqn8} and the only role of the stiffness $E$ is to set the 
stress scale. This means that the behavior of the stress components and 
and sample size as a function of $r$ is independent of the local 
force law. In particular, in the limit of infinitely rigid particles, the 
same results remain true up to a stress scale which may be fixed through a 
confining pressure. More generally, both the local interactions and 
the mass or heat transfer influence the stress scale.       

By analogy with molecular solids, we  
introduce a ``theoretical'' tensile strength $\sigma^y_{th}$ based on the 
interactions between two particles \cite{herrmann90}. 
According to Eq. \ref{eqn3}, the orthoradial stress is 
$\sigma_{\theta\theta} = n_c \  \langle f_\theta \ell_\theta \rangle 
\simeq n_c \langle \ell \rangle \ \langle f_\theta \rangle$, where $n_c$ is the 
density of contacts and $\langle \ldots \rangle$ designs averaging over the control 
volume.  The largest value of $\sigma_{\theta\theta}$ 
in tension corresponds to the limit  
where all forces are polarized in the same direction and 
they have all reached the largest tensile force $f_n^y$. This defines the 
``theoretical'' tensile strength
\begin{equation}
\sigma^y_{th} = n_c    \langle \ell \rangle   f_n^y  
\label{eqn10}
\end{equation}

In our simulations, the measured tensile strength  $ \sigma^y$ 
is by a factor $\simeq 4.3$ below $\sigma^y_{th}$. In molecular solids, 
a similar discrepancy between $\sigma^y_{th}$, defined from atomic interactions in 
a regular atomic arrangement, and $ \sigma^y$  
stems from ``built-in'' disorder at different scales leading to stress 
concentration. In a granular solid, the  
disorder is ``intrinsic'' to the structure. As a result, 
the contact forces 
both in cohesive and noncohesive granular media 
have a wide distribution with a decreasing exponential 
shape for strong forces \cite{radjai98b,radjai00a}. It seems that 
$ \sigma^y$ reflects the {\em strongest} contact force at failure, 
whereas $\sigma^y_{th}$ represents the 
{\em mean} tensile force by construction. Indeed, 
the strongest tensile force 
in simulations is by a factor $\simeq 5$ below the mean tensile force, 
and this is  close to $\sigma^y_{th} / \sigma^y \simeq 4.3 $. 
As far as we know, this is the first example showing how  
the local force inhomogeneities in a granular material 
control a macroscopic property, namely the tensile strength.

In summary, 
our numerical data and their comparison with 
an analytical evaluation of stresses in the elastic 
domain and at failure suggest that a macroscopically  elastic behavior is 
relevant up to crack initiation, as in molecular solids. 
However, the tensile strength is dependent on the inhomogeneous 
transmission of forces. 
The simple test described in this Letter not only 
provides reproducible results, but it  
has also the advantage of combining       
features of discrete modeling with theoretical predictability at the 
macroscopic scale. 

This approach may now be used to investigate and to predict 
the tensile thresholds and crack propagation in 
cohesive granular materials as a function of the intial density and 
anisotropy of the material or the possible couplings of the 
local cohesion with mass and heat transfer in the pores as 
in fine soils and granular rocks \cite{tarbuck02}. 
The theoretical approach can be extended to other structured media involving 
mesoscopic length scales, such as gels \cite{mrani95}, 
cellular media \cite{schwarz02}, layered structures such 
as wood \cite{kubler87} and pastes\cite{ponsart03}. 
Swelling or shrinkage may occur 
as a result of cellular growth (in biological systems) 
or the evolution of local variables such as water content and temperature.  

\vspace*{0.5cm}
It is a pleasure to thank J.-C. B\'enet and J. N. Roux for helpful suggestions.

\bibliographystyle{apalike}

\bibliography{granular}

\end{document}